# Resonant THz detection by periodic multi-gate plasmonic FETs

Yuhui Zhang and Michael S. Shur

*Abstract*—We show that a periodic multi-grated-gate structure can be applied to THz plasmonic FETs (TeraFETs) to improve the THz detection sensitivity. The introduction of spatial non-uniformity by separated gate sections creates regions with distinct carrier concentrations and velocities, giving rise to harmonic behaviors. The resulting frequency spectrum of DC voltage response is composed of "enhanced" and "suppressed" regions. In the enhanced region, the amplitude of response voltage can be enlarged up to ~100% compared to that in a uniform channel device. The distribution pattern of those regions is directly related to the number of gate sections ($N_s$). A mapping of response amplitude in an $N_s$-frequency scale is created, which helps distinguish enhanced/suppressed regions and locate optimal operating parameters.

## I. Introduction

Short channel field-effect transistor (FET) operated in plasmonic regime at sub-THz or THz frequencies (often referred to as TeraFETs [1, 2]), are promising devices for THz applications such as sensing [3-6], imaging [7-9], and beyond-5G communication [1, 3]. TeraFETs can work in the plasmonic resonant (ballistic or viscous) regimes [10, 11], in which the plasma waves are generated [12, 13]. Such hydrodynamic-like property allows TeraFETs to break the frequency limitation set for collision-dominated devices and operate at GHz to THz ranges. TeraFETs are also tunable by the gate bias or doping or illumination [14-16], The high speed of plasma waves enables TeraFETs to be a strong candidate for ultrashort pulse detection [17, 18].

To facilitate the industrial applications of TeraFETs, one of the key issues is to improve the detection sensitivity. As was discussed in [1], further improvement in the noise-equivalent power of TeraFETs is required to enable 6G communication applications. A straightforward way is to use better materials, e.g. materials with high mobility ($\mu$) and high effective mass ($m^*$), so as to elevate the device quality factor ($Q = \omega_p\tau$, where $\omega_p$ is the plasma frequency, $\tau = \mu m^*/e$ is the momentum relaxation time) [19]. We have demonstrated that p-diamond could be a valid candidate for high-sensitivity THz and sub-THz detections [20-22]. In addition to the material consideration, one can also resort to new physical/structural designs. The non-uniform structures, such as grating gates [23-27], dense arrays [28-30], and plasmonic crystals [23, 31], were

The authors are with the Department of Electrical, Computer and Systems Engineering, Rensselaer Polytechnic Institute, Troy, NY 12180 USA (e-mail: shurm@rpi.edu, zhangy79@rpi.edu).

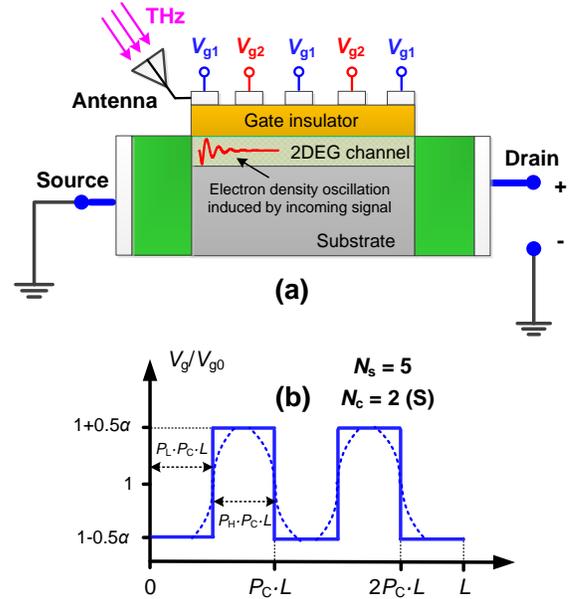

**Fig. 1.** (a) Schematic of THz detection by a periodic multi-gate TeraFET. (b) the resulting spatial distribution of DC gate bias. The ideal and realistic distribution curves are illustrated in solid and dashed lines, respectively. Here $N_s$ is the number of split gates, $P_L$ and $P_H$ are the duty ratios of high and low voltage in one high-low cycle, respectively. $P_C$ is the ratio of one high-low cycle in the whole channel. $N_c$ is the number of complete high-low cycles. We define $P_L + P_H = 1$, $(N_C+P_L)P_C = (N_C+1-P_H)P_C = 1$. Besides, $\alpha$ is a voltage modulation factor, $V_{g0}$ is a reference gate voltage.

introduced and proved to be effective in improving the TeraFET detection performance.

The introduction of specifically-arranged non-uniform structures in TeraFETs can modify the carrier density, static field distribution, and plasma wave velocity along the device channel, thus altering the THz rectification properties and/or the wave propagation features. For example, with a split-gate structure and a graded doping (i.e. the grating-gate), the circularly polarized THz radiation can be rectified by the TeraFET, inducing DC currents in both parallel and transverse directions [15, 27]. It was shown that the DC current flux in the transverse direction is related to the helicity of the THz radiation, and this current is dramatically enhanced near the plasmon resonant frequencies. The multi-gates can also be rearranged to create a concatenated FETs dense array, where the source, drain, and gate are all split into fingers and nested together to form the repeated unit cells [29, 30, 32]. Such short-period grating of metal contacts strengthens the device asymmetry and serves as an effective antenna coupling incident

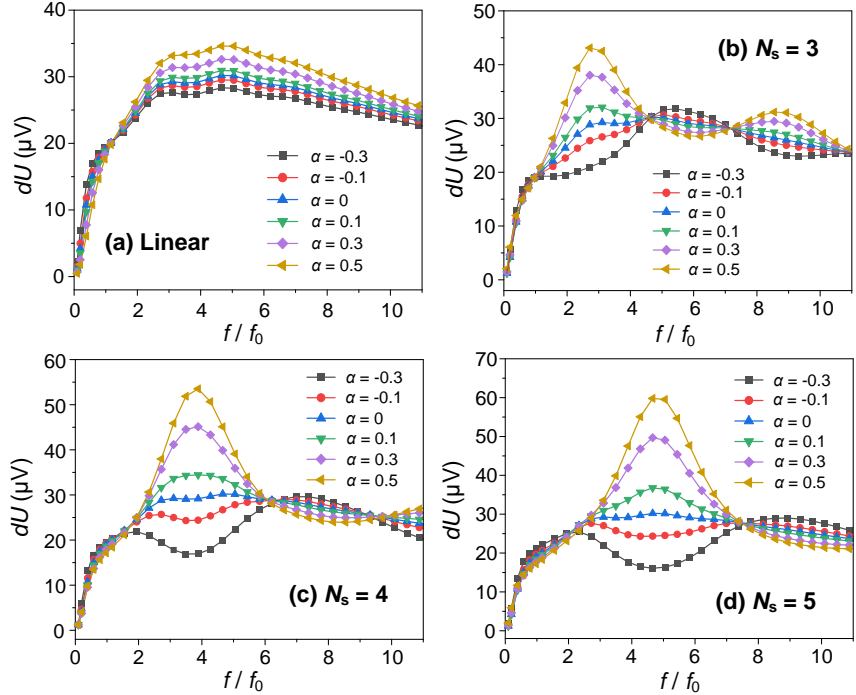

**Fig. 2.** DC response voltage ($dU$) as a function of frequency ($f$) in a Si periodic multi-gate TeraFET with different values of $N_s$. Here $f_0$ is the fundamental resonant frequency and $f_0 \approx 0.515$ THz. Other parameters: $V_{g0}$ = -0.2, $V_{th}$ = 0.2 V, $V_{am}$ = 2 mV. $T$ (temperature) = 300 K, assume thermal equilibrium.

THz radiations, thereby improving the detection sensitivity. In addition, the grating-gate structure can also synergize with the applied DC current to create full transparency and the amplification of THz radiation [33].

In our recent work [2], we used a spatially non-uniform gate capacitance or threshold voltage to induce the channel nonuniformity. Those structures could modify the transport properties of plasma waves and enhance or suppress the non-resonant photoresponse [2, 29].

In this work, we will discuss the effects of periodic multi-gate structures on the resonant THz detection performance in a wide spectral range that includes several plasmonic modes. As will be shown later, the periodic multi-gate TeraFETs possess strong harmonic behaviors and can achieve a ~100% improvement in DC voltage response near the resonant peaks.

## II. MODEL AND EQUATIONS

In this work, we consider a periodic multi-gate TeraFET structure to achieve high-sensitivity resonant THz detection. Fig. 1(a) shows the schematic of the structure. The gates are driven by periodic-in-space DC excitations. Compared to the varying capacitance or varying threshold voltage design in our previous work [2], this periodic gate structure could be easier to fabricate. The number of gate sections ($N_s$) is adjustable. With the repetitive excitation of DC biases $V_{g1}$ and $V_{g2}$, the spatial distribution of DC gate voltage can be approximated by a square-wave voltage shown in Fig. 1(b). This approximation can be verified via electrostatic modeling (see supplementary material). A more realistic consideration is to include the transition regions between each two adjacent sections, as illustrated by dashed lines in Fig. 1(b). The transition region here results from the separation (i.e. the ungated region) between two adjacent gate segments. We assume that the length of the separated region is short so that the carriers underneath can be screened by the peripheral voltage of neighboring gates. Therefore, we still consider the transition regions as gated regions.

We use a 1D hydrodynamic model [11, 14, 34] to simulate the response of the proposed TeraFET structure. The detailed introduction and validation of the model can be found in [11]. The key equations are:

$$\frac{\partial n}{\partial t} + \nabla \cdot (n\boldsymbol{u}) = 0 \qquad (1)$$

$$\frac{\partial \boldsymbol{u}}{\partial t} + (\boldsymbol{u} \cdot \nabla)\boldsymbol{u} + \frac{e}{m^*}\nabla U + \frac{\boldsymbol{u}}{\tau} - \nu\nabla^2\boldsymbol{u} = 0 \qquad (2)$$

where $n$, $\boldsymbol{u}$ are the carrier density and hydrodynamic velocity, respectively. $m^*$ is the effective mass of carriers. $U$ is the gate-to-channel voltage defined as $U(x) = U_0(x) - U_{ch}(x)$, where $U_0(x) = V_g(x) - V_{th}(x)$ is the gate bias beyond threshold, $U_{ch}(x)$ is the channel potential. A unified charge-control model [35, 36] is used to related $n$ and $U$:

$$n(U) = \frac{C_g \eta V_t}{e} \ln(1 + \exp(\frac{U}{\eta V_t})) \qquad (3)$$

where $V_t = k_B T/e$ is the thermal voltage ($k_B$: Boltzmann constant, $T$: temperature, fixed at 300 K). $\eta$ is an ideality factor.

In this work, we focus on the effects of non-uniform $V_g$ on the detection performance in the absence of any helicity-sensitive effects. We consider the external THz radiation shining onto the leftmost gate section, as shown in Fig. 1(a). The boundary condition at the source side can now be approximated by $U(0,t) = U_0(0) + U_a(0,t)$ [19], where $U_a(0,t) = V_{am} \cdot \cos(\omega t)$ represents the AC small-signal voltage induced by the incoming THz radiation. On the drain side, an open circuit condition is used, i.e. $J(L, t) = 0$, where $J$ is the current flux

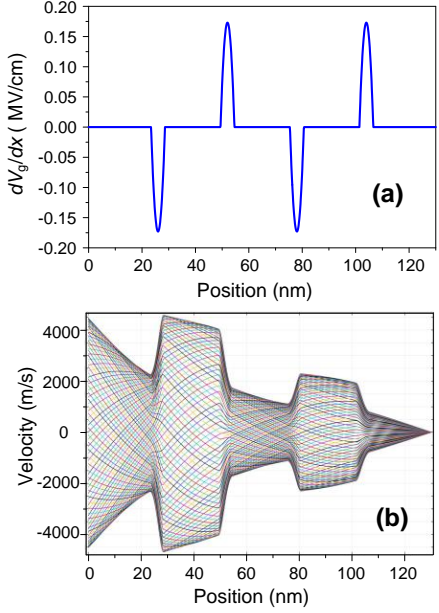

Fig. 3. Spatial distributions of (a) the gradient of DC gate bias ($dV_g/dx$) and (b) variation contour of carrier velocity at $\alpha = 0.3$. Each curve in (b) represents the carrier velocity distribution $u(x)$ at a given moment in one AC period, and 50 consecutive moments are included. Other parameters follow those in Fig. 2.

density, $L$ is the channel length.

## III. RESULTS AND DISCUSSION

### A. Frequency dependent profiles

Based on the above model settings, we simulate our device and evaluate the frequency spectrum of DC source-to-drain response voltage ($dU$). $dU$ is proportional to the intensity of THz signal, which, in turn, is proportional to the squared THz voltage amplitude. For a single gate section, the response has the form [19]

$$dU = \frac{eV_{am}^2}{4m^*s} f(\omega) \quad (4)$$

where $\omega$ is the angular driving frequency, $f(\omega)$ is a frequency-dependent function associated with the plasma wave (or damped electron wave) propagation properties.

The results under linear $V_g(x)$ profile and 3 different $N_s$ values under multi-gate structure are presented in Fig. 2. These results are for a Si FET with 50% duty ratio ($P_L = P_H = 50\%$, see Fig. 1), and $V_{g0} = -0.2$ V, $V_{th} = 0.2$ V. Thus the device is driven into the subthreshold mode with a relatively large voltage response [37, 38]. To improve convergence, the continuous-first-derivative transition regions are set between neighboring gate sections, and the relatively size of those regions ($T_z$, the ratio of total transition region size over the whole channel size) is fixed at $T_z = 0.1$ (see more details in supplementary materials). The number of sections varied from 2 to 7. The voltage applied to different gates varied (see Fig. 1(b)).

Fig. 2(a) shows the result under a linear varying gate voltage: $V_g(x) = V_{g0}(1+\alpha(x-0.5L)/L)$. We can see that with the increase of $\alpha$ (or the decrease of DC gate bias swing from source to drain since $V_{g0}$ is negative), $dU$ decreases when $f < f_0$, where $f_0 = s/4L$ is the fundamental resonant frequency, $s$ is the plasma wave velocity [21, 37]. This region corresponds to the non-resonant operation region of the device. Using the methods in [2] and [37], we can get the expression of DC response in this region (see supplemental material for detailed derivations):

$$dU = \frac{eV_{am}^2}{4mS_e^2}(1+\beta - \frac{1+\beta\cos(2k_r L)}{\cosh(k_1 L)\cosh(k_2 L)}) \quad (6)$$

where $\beta = 1/\sqrt{1+(\omega\tau)^{-2}}$, $k_1, k_2$ are wave vectors of the plasma wave:

$$k_1 = \frac{\alpha_1^*}{2}(\frac{S_0}{S})^2 + \sqrt{(\frac{\alpha_1^*}{2}(\frac{S_0}{S})^2)^2 + ik_0^2} \quad (7)$$

$$k_2 = \frac{\alpha_1^*}{2}(\frac{S_0}{S})^2 + \sqrt{(\frac{\alpha_1^*}{2}(\frac{S_0}{S})^2)^2 - ik_0^2}$$

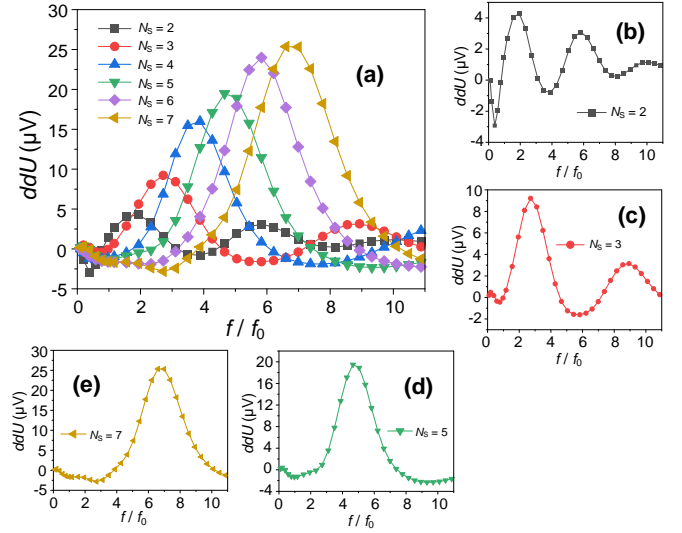

Fig. 4. (a) $ddU$ as a function of $f/f_0$ for $N_s = 2\sim 7$ and (b)-(d) separated plots with $N_s = 2, 3, 5$, and 7, respectively. Here $ddU$ is defined as $ddU = dU(\alpha=0.3) - dU(\alpha=0)$. Other parameters follow those in Fig. 2.

Here $\alpha_1^* = \frac{1}{|V_{g0}|}\frac{\partial V_g}{\partial x}, S_0 = \sqrt{\frac{e|V_{g0}|}{m^*}}, S = \sqrt{\frac{\eta eV_t}{m^*}(1+\exp(-\frac{U_0}{\eta V_t}))\ln(1+\exp(\frac{U_0}{\eta V_t}))}$.

Besides, $k_0 = k_0 = (\omega/S^2\tau)^{0.5}$ is the wave vector in the uniform channel ($\alpha_1^*=0$), $k_r$ is the real part of $k_1$ or $k_2$. A transition of variation trend with respect to $\alpha$ occurs at around $f = f_0$. Beyond $f_0$, the plasmonic resonance can be achieved, and $dU$ decreases with increasing $\alpha$. Now the response curve does not follow equation (6). Within $\alpha \in [0, 0.5]$, the maximum improvement of $dU$ is around 20%. Those results agree with our observations of linearly varying gate capacitance or threshold voltage [2].

Fig. 2(b) shows the result of $dU$ vs $\alpha$ under $N_s = 3$. Compared to Fig. 2(a), the 3-segment multi-gate TeraFET exhibits a distinct response profile. As $f$ rises, the response voltage oscillates, and the variation trend of $dU$ with respect to $\alpha$ changes multiple times. If we define the regions where $dU$ increases with rising $\alpha$ as the "enhanced" regions, and the regions where $dU$ decreases with rising $\alpha$ as the "suppressed" regions, we can see that the enhanced and the suppressed regions appear alternatively with the increase of frequency. More interestingly, the positions of those regions are directly related to the number of gate sections. For example, the peak response voltage in the first enhanced region (which is also the peak $dU$ in the whole frequency range) is at $f = 3f_0$, the position of response valley in the first suppressed region is $f = 6f_0$, and the position of peak response in the second enhanced region is at $f = 9f_0$. Thus, we conclude that the frequency at which the maximum response is reached is around

$$f_{pmax} = N_s f_0 \quad (8)$$

and the frequency gap between two adjacent peaks or valleys is

$$df_p = 2N_s f_0 \quad (9)$$

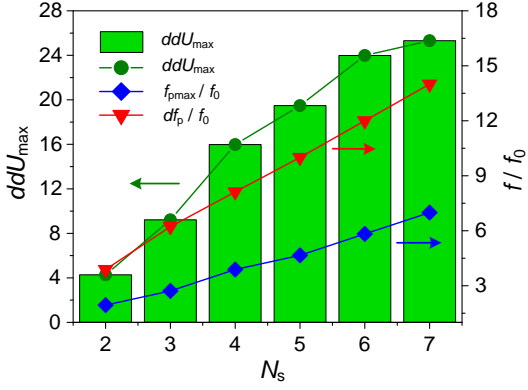

**Fig. 5.** The peak value of $ddU$ ($ddU_{max}$), the frequency at which $ddU$ reaches the maximum ($f_{pmax}$), and the frequency gap between two adjacent peaks or valleys ($df_p$) as functions of $N_s$. Other parameters follow those in Fig. 2.

This result is similar to the one reported in [33], where the THz transmission spectra is controlled by the gate separations in a grating-gating graphene FET, and the resonant frequency is determined by the unit finger gate width ($\sim L/N_s$). With the modulation of two DC bias in our work, more harmonic behaviors can be observed, as will be discussed later.

Equation (8) and (9) can be further verified by the simulations under other $N_s$ values. For example, in Fig. 2(c) where $N_s = 4$, the peak frequency is at $4f_0$ and the distance between two adjacent peaks or two adjacent valleys are $8f_0$. In Fig. 2(d) where $N_s = 5$, the values of $f_p$ and $df_p$ are $5f_0$ and $10f_0$, respectively. Also, a 100% increase in $dU$ (compared to the uniform channel case) is achieved when $\alpha$ reaches 0.5. Note that the peaks and valleys are not located at the fundamental resonant frequency, but at the higher order harmonics. Therefore, the introduction of multiple gate sections activated the harmonic components in the system, resulting in the distribution of enhanced and suppressed regions. The underlying mechanism could be related to the reflection of plasma waves or carrier drift between neighboring sections due to the carrier concentration barriers. Those reflections change the wave propagation properties (i.e. $k_1$ and $k_2$) and shorten the effective channel length, thereby leading to the excitation of harmonic peaks and valleys. Fig. 3(a) shows the spatial distribution of gate-induced field ($dV_g/dx$) along the channel. The abrupt change of DC gate bias in the narrow transition regions creates a large field on the order of 0.1 MV/cm. The electrons passing the transition regions get accelerated or de-accelerated, forming the separated velocity distribution regions, as demonstrated in Fig. 3(b), and possibly induces the reflections of plasma waves in between.

The above harmonic excitation mechanism can be seen as a result of abrupt changes in channel properties, as opposed to the gradual changes reported in our previous work [2]. In a gradually varying channel, the response performance is related to the changing rate of channel parameters (e.g. the gate capacitance, threshold voltage, DC gate bias). While in multi-gate setup, we can verify from simulation that the response $dU$ is insensitive to the transition region size $T_z$ (see supplementary material). This indicates that the response profile is now level-sensitive, as opposed to the gradient-sensitive ones in [2]. Therefore, the analytical approaches developed in [2] can no longer be applied here.

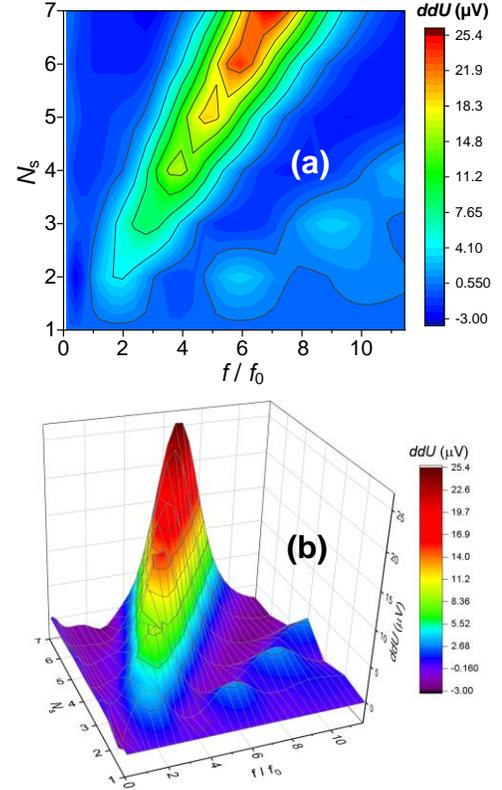

**Fig. 6.** 2D mappings of $ddU$ in a $N_s$-$f/f_0$ scale. (a) 2D contour plot, (b) 3D colormap surface plot. The data presented are the same as those in Fig. 5.

To further investigate the variation trend of $dU$ with frequency, we define a differential response voltage $ddU = dU(\alpha=0.3) - dU(\alpha=0)$, and plot its frequency profile at different $N_s$ values, as shown in Fig. 4. Here $ddU$ signifies the net enhancement or suppression of $dU$ at $\alpha = 0.3$ as compared to the uniform channel case. In Fig. 4(a), the amplitude of $ddU$ rises with the increase of $N_s$. This suggests that the enhancement effect strengthens as the channel becomes more non-uniform. With the rise of frequency, $ddU$ oscillates and exhibits multiple peaks and valleys, as shown in the separated plots Fig. 4(b)~Fig. 4(d). For quantitative analysis, we plot Fig. 5 where $ddU_{max}$, $f_{pmax}$ and $df_p$ as functions of $N_s$ are presented. One can check that the $f_{pmax}$ and $df_p$ curves follow Equation (8) and (9). The $ddU_{max}$ increases with the rise of $N_s$, but a saturation trend is observed when $N_s$ becomes large. This saturation could be related to the change of wave reflection characteristics as the length of each gate section shortens, which sets a limit to the maximum improvement of $dU$.

*B. Mapping of enhanced/suppressed regions*

To better understand how the response changes with frequency and gate structure, we create a map of $ddU$ in a $N_s$-$f/f_0$ scale, as shown in Fig. 6. In the map, the enhanced regions are exhibited as "mountains" while the suppressed regions are presented as "valleys" - a result of the present $ddU$ definition. The highest mountain group is located at $f = N_s f_0$, as shown in Fig. 6(a), which corresponds to the maximum (the first) resonant peak in each case. The second mountain series are at $f = 3N_s f_0$, demonstrating the secondary resonant peaks. Between these two mountain groups is a valley group located at $f = 2N_s f_0$. In general, the mountain clusters can be expressed by $f = (2n+1)N_s f_0$, where $n = 0,1,2…$, and the valley clusters follow $f = 2nN_s f_0$.

Fig. 6(b) shows the direct comparison of the heights of different mountains (i.e. the amplitudes of response peaks). Clearly, the mountain height in each group increases with the increase of $N_s$, and the average/maximum height in the first mountain group is much larger than that in the second mountain group. Thus, to achieve a high response, the TeraFET should operate in the first mountain group, and in general a large gate section number is preferred.

### C. Limits of response tuneability

The results in section III.A and section III.B demonstrate that adopting periodic multi-gate structure in TeraFETs can effectively alter the DC voltage response and achieve over ~100% improvement in $dU$ at certain frequencies. The amplitude of $dU$ can be tuned by $N_s$ and $\alpha$. In general, a larger $N_s$ or $\alpha$ leads to a higher responsivity in the enhanced region, but the values of $N_s$ or $\alpha$ cannot grow infinitely due to several built-in limits. Here we discuss those limits.

(1) Breakdown voltage (vertical). To prev ent the breakdown of the barrier material, the following is required

$$\frac{(1+0.5\alpha)|V_{g0}|}{d_b} < E_b \rightarrow \alpha < 2(\frac{E_b d_b}{|V_{g0}|} - 1) \quad (8)$$

For example, if $E_b = 3$ V/nm, $d_b = 4$ nm, $|V_{g0}| = 0.2$ V, we get **$\alpha < 46$**.

(2) Breakdown voltage (transverse). Let $D$ denotes the transition region length between two gate sections, and $D$ is related to $T_z$. To prevent dielectric breakdown in the transition region, we need

$$E_b > \frac{(V_{g1} - V_{g2})}{D} = \frac{\alpha|V_{g0}|}{D} \rightarrow \alpha < \frac{E_b D}{|V_{g0}|} \quad (9)$$

If $E_b = 3$ V/nm, $|V_{g0}| = 0.2$ V, $D = 2$ nm, we get **$\alpha < 30$**.

(3) Conductivity limit. When the gate bias decreases in the subthreshold region, the carrier concentration can reduce to very low, so as to choke the current conduction. Assume that the minimum conductivity required for sustaining current conduction is $\sigma_{cr} = e\mu n_{cr}$, where $n_{cr}$ is the critical carrier density. Using Equation (3), we get:

$$n_{cr} = \frac{\sigma_{cr}}{e\mu} \leq \frac{C_g \eta V_t}{e} \ln(1 + \exp(\frac{V_{g0}(1+0.5\alpha)}{\eta V_t})) \quad (10)$$

$$\rightarrow \alpha \leq 2(\frac{\eta V_t}{V_{g0}} \ln(\exp(\frac{en_{cr}}{\eta V_t}) - 1) - 1) \approx 2(\frac{\eta V_t}{V_{g0}} \ln(\frac{en_{cr}}{\eta V_t}) - 1)$$

If $n_{cr} = 10^{14}$ m$^{-3}$, $V_{g0} = -0.2$ V, $V_t = 0.026$ V ($T = 300$ K), $\eta = 2$, we get **$\alpha < 2.2$**.

(4) Fabrication limit. The fabrication lab conditions determine the maximum number of separated gates that can be built in a TeraFET. If the minimum achievable size is $L_{min}$, then we get $N_{s-max} = [L/L_{min}]$, where $[k]$ denotes the nearest integer that does not exceed $k$. For example, with $L = 250$ nm, $L_{min} = 65$ nm, we get **$N_{s-max} = 3$**.

The above conditions, along with other more delicate mechanisms (e.g. the self-capacitance and the built-in voltage between two adjacent gate sections), set limit to the tuning of $dU$ in periodic multi-gate TeraFETs. Despite all those constraints, an ~100% improvement can still be achieved near the maximum resonant peak, as demonstrated in Fig. 2(d).

## IV. Conclusion

When a periodic multi-gate structure is applied in TeraFETs, the resonant THz detection performance can be improved. The hydrodynamic simulation showed that in periodic multi-gate TeraFETs, the harmonic response peaks were excited, and thus the DC response voltage $dU$ near the harmonic frequencies could increase ("enhanced") or decrease ("suppressed") compared to $dU$ in the uniform-channel TeraFETs. The excitation of harmonics peaks could be related to the strong gate-induced field in the transition regions, which accelerates or de-accelerates the carriers and possibly leads to the reflection of plasma waves on the boundaries of gate sections. The frequency spectrum of $dU$ was separated by the "enhanced" and "suppressed" regions, and the distribution of those regions was related to the number of gate splits. The maximum improvement on $dU$ reached beyond 100%. The tunability of $dU$ via gate parameters is limited by the breakdown voltage, conductivity, fabrication resolution, and other more delicate effects. A mapping of variation in $dU$ helps distinguish enhanced/suppressed regions and locate optimal operating parameters.